\documentclass[%
reprint,showkeys,
nofootinbib,  
amsmath,amssymb,
aps,
]{revtex4-2}

\usepackage{graphicx}
\usepackage{dcolumn}
\usepackage{bm}
\usepackage{siunitx}
\usepackage{xcolor}
\usepackage{balance}
\usepackage[english]{babel}
\DeclareMathOperator{\Tr}{Tr}

\begin{document}
	
	\preprint{APS/123-QED}
	
	\title{Swelling induced debonding of thin hydrogel films grafted on silicon substrates}	
	\author{Anusree Augustine}
	\author{Yvette Tran}%
	\author{Emilie Verneuil}%
	\author{Antoine Chateauminois}%
	\email{antoine.chateauminois@espci.fr}
	\affiliation{%
		Soft Matter Science and Engineering Laboratory (SIMM), CNRS UMR 7615, ESPCI Paris, PSL University, Sorbonne Universit\'e, F-75005 Paris, France
       \textbackslash\textbackslash
	}%
		
	\author{Nicolas Gauthier}
	\author{Marc Veillerot}
	\affiliation{
		University Grenoble Alpes, CEA, LETI, F-38000 Grenoble, France\\
	}%
	\author{Bangguo Zhu}
	\author{Chung-Yuen Hui}
	\altaffiliation[Also at ]{Global Station for Soft Matter, GI-CoRE, Hokkaido University, Sapporo, Japan}
	\affiliation{%
	Department of Mechanical and Aerospace Engineering, Cornell University, Ithaca, NY 14853, USA
 \textbackslash\textbackslash
	}%
	
	
	\date{\today}
	
	\begin{abstract}
		We report on the delamination of thin ($\approx \mu$m) hydrogel films grafted to silicon substrates under the action of swelling stresses. Poly(dimetylacrylamide) (PDMA) films are synthesized by simultaneously cross-linking and grafting preformed polymer chains onto the silicon substrate using a thiol-ene reaction. The grafting density at the film/substrate interface is tuned by varying the surface density of reactive thiol-silane groups on the silicon substrate. Delamination of the films from well controlled line defects with low adhesion is monitored under a humid water vapor flow ensuring full saturation of the polymer network. A propagating delamination of the film is observed under the action of differential swelling stresses at the debonding front. A threshold thickness for the onset of this delamination is evidenced which is increasing with grafting density while the debonding velocity is also observed to decrease with an increase in grafting density. These observations are discussed within the framework of a nonlinear fracture mechanics model which assumes that the driving force for crack propagation is the difference between the swelling state of the bonded and delaminated parts of the film. Using this model, the threshold energy for crack initiation was determined from the measured threshold thickness and discussed in relation to the surface density of reactive thiol groups on the substrate.		
	\end{abstract}
	
	\maketitle

\section*{Introduction}
Hydrogel coatings are suitable candidates for many applications in the biomedical and engineering fields where properties such as bio-compatibility, lubrication, hydrophilicity, and transparency are desired. Indeed these hydrophilic polymer networks may absorb water by swelling to several times their dry thickness. As such, hydrogel coatings are, for example, possibly good candidates for anti-fog applications, as proposed by several teams \cite{grube2015,chevallier2011,shibraen2016,park2016} who discussed their efficiency in terms of light transmission while submitted to warm humid air. Indeed, the coating should act as a reservoir for humidity and observation is made \cite{shibraen2016,park2016,delavoipiere2018} that the higher the film thickness, the more water is absorbed and the later mist appears. However, the viability of such anti-fog coatings is conditioned by their ability to sustain stresses without damage.\\
Inherent to the coating geometry is the very large aspect ratio which results in the development of internal stresses upon swelling of a coating bound to a rigid substrate. Indeed, swelling occurs along the thickness direction while compressive stresses build up in the constrained in-plane direction. For hydrogel films strongly anchored to the substrate, these swelling stresses were observed to induce instabilities on the free surface in the form of creases.~\cite{Trujillo2008,tanaka1987,yoon2010,xu2015} In situations where the film is not strongly bound to the substrate, swelling stresses can alternately result in delamination followed by large scale folding and blistering of the hydrogel films. Such mechanisms have been reported by Velankar~\textit{et al.}~\cite{Velankar2012} for poly(dimethylsiloxane) films swollen with toluene causing growth of sharp folds of high aspect ratio. Other morphologies in the shape of craters were also reported by Sharp~\textit{et al.}~\cite{Sharp2002} for water immersed ultrathin films of poly(d,l-lactide) as a result of a blistering process which occurs when the films delaminate.\\
In a morphogenetic perspective, controlled debonding and blistering of swollen hydrogel films on patterned substrates were also envisaged as a tool to generate switchable actuators or micro/nanofluidic devices. As an example, Xu and coworkers~\cite{xu2016} elaborated poly(\textit{N}-isopropylacrylamide)-\textit{co}-sodium acrylate) (pNIPAM) gel layers on micro-patterned electrode surfaces which allowed to trigger electrochemically reversible delamination and blistering of the films. Similarly, Takahashi~\textit{et al.}~\cite{Takahashi2019} explored in a systematic way the 3D architectures which are generated from the swelling-induced buckling instabilities within hydrogel films on patterned silanized substrates. In these studies, the control of the architectures relies on the absence of any debonding at the film/substrate interface outside the patterned zones. \\
In a different perspective, here we take advantage of patterning techniques to promote swelling-induced delamination of thin ($\approx \si{\micro\meter}$) hydrogel films grafted onto silicon substrate from well controlled nucleation sites with low adhesion. Using this methodology, we offer to quantitatively relate the delamination of swollen hydrogel films to their chemical bonding to a rigid substrate that we vary using well-controlled grafting densities.\\
We first focus on the analysis of the debonding and blistering mechanisms which result from the swelling of the films under a water vapor flow ensuring full saturation of the hydrogel network. In rigid film/substrate systems experiencing stress mismatch, blistering and delamination are often considered to be driven by the bending stresses induced within the buckled film (see \textit{e.g.} references~\cite{gioia1997,hutchinson1992} for a review). For soft hydrogel films bonded to a rigid substrate, we anticipate that the driving mechanisms can be different owning to the comparatively higher compressive stresses and lower elastic modulus. Indeed, we show here that for soft hydrogel films bonded onto rigid substrates, delamination is instead triggered by localized differential stresses which arise at the debonding front from the difference in the swelling state in the bonded and delaminated parts of the film. We also show that debonding only occurs above a thickness threshold which depends on the grafting density.\\
As theoretically investigated by Gent~\textit{et al.}~\cite{gent2001} for rubber layers, the energy released when a swollen layer separates from a substrate and swells further can indeed be quite high and justify the occurrence of spontaneous debonding. Here, we develop a non linear fracture mechanics analysis of the threshold energy required for crack extension at the subtrate/film interface under the action of swelling stresses. Using this theory, we determine how the threshold energy release rate for debonding is varying as a function of the grafting density from the experimental threshold thickness.\\
%
\section*{Experimental section}
%
\subsection*{Synthesis of grafted PDMA films}
\begin{figure} [!ht]
	\centering
	\includegraphics[width=1\linewidth]{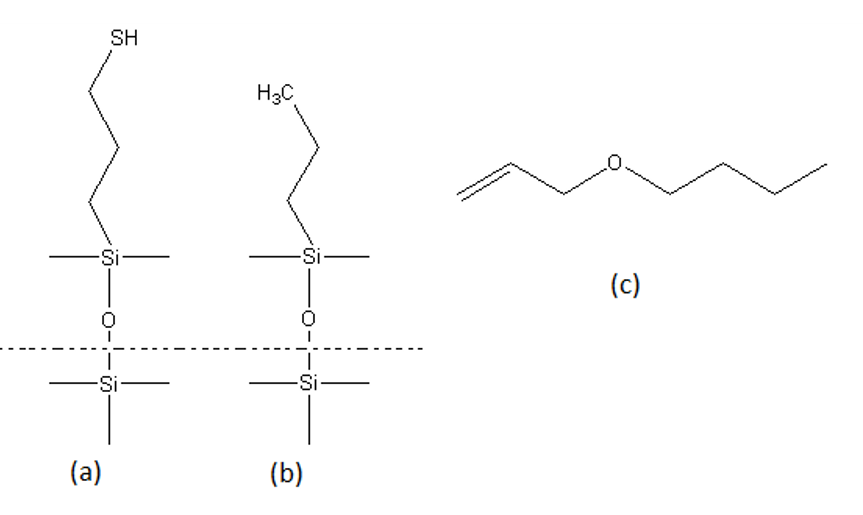}
	\caption{Silicon wafer grafted with (a) (3-mercaptopropyl)trimethoxysilane and (b) propylsilane. (c) Allyl butyl ether used to passivate the grafted thiol groups.}
	\label{fig:silane}
\end{figure}

Poly(dimethylacrylamide) hydrogel films are synthesized by simultaneously crosslinking and grafting preformed polymer chains onto silicon wafer substrates with a thiol-ene click reaction. As fully described elsewhere~\cite{chollet2016a,li2015}, ene-reactive PDMA is first synthesized by free radical polymerization of dimethylacrylamide and acrylic acid which is then modified by amidification using allylamine. The ene-functionalized PDMA is subsequently purified by dialysis against water and recovered by freeze-drying.\\
The adhesion between the hydrogel film and the wafer is achieved through thiol-modification of the silicon surface (Fig.~\ref{fig:silane}a). In order to vary the interface strength, we tuned the grafting density of the gel onto the silicon wafer by playing with the density of active thiol-silane groups at the silicon wafer surface. This goal is achieved using a mixture of \textit{n}-propyltrimethoxysilane  and 3-mercaptopropyl trimethoxysilane for the silanization of the silicon wafer (fig.~\ref{fig:silane}a,b).
%
While mercaptosilane has a thiol group at one end which can react with the ene-functionnalized PDMA chains to form covalent bonds at the interface, propylsilane is not reactive. Here, 50~vol\% and 100~vol\%  mercaptosilane in volume is used in the silane mixture in order to vary the density of grafted thiol groups. The silicon substrate is first cleaned in a UV ozone chamber and quickly transferred into a sealed reactor filled with nitrogen where a solution of 30\%vol of silane in dry toluene is introduced. After 3 hours soaking in the silane solution under nitrogen, the silicon wafer is rinsed and sonicated in toluene before drying. Using a combinaison of X-ray Photoelectron Spectroscopy (XPS) and Time of Flight Secondary Ion Mass Spectrometry (ToF SIMS) analysis detailed elsewhere~\cite{augustine2023}, the areal density of grafted thiol groups was found to vary non linearly from 0.2 to 0.6 per $\si{\square\nano\meter}$ when the percentage of mercaptosilane in the reactive silane mixture is increased from 50~vol\% to 100~vol\%.\\
Prior to spin-coating on the thiol-modified silicon wafers, the ene-functionalized PDMA and dithioerythritol crosslinker are dissolved in dimethylformamide (DMF). In order to vary film thickness, spin coating was performed at 3000~rpm using solutions with different polymer concentrations. After spin-coating, polymer films were annealed at 150~$^\circ C$ for 24 hours under vacuum to activate the thiol-ene reaction. This procedure resulted in PDMA films with thicknesses $h$ ranging from 100~$\si{\nano\meter}$ to 1.2~$\si{\micro\meter}$ in the dry state, as measured by ellipsometry. All the film thicknesses $h$ reported in this paper refer to the dry state. Unless otherwise stated, all the results reported in this study were obtained with films with a swelling ratio of $2.7 \pm 0.3$, as measured by additional ellipsometry measurements with the films fully swollen in water. This value is found to be independent on the grafting density and on film thickness within the investigated range~\cite{augustine2023}. Such a constant swelling ratio ensures a constant swelling stress within the films at all thicknesses.\\
\subsection*{Patterning of the interface with strip defects}
Debonding mechanisms were promoted from well controlled line defects on the silicon substrate where the strength of the interface is reduced as a result of localized passivation of the reactive thiol groups grafted on the silicon substrate. Such a passivation is achieved by covalent bonding of the thiol-silane goups on the silicon wafer with allyl butyl ether as a passivator (fig.~\ref{fig:silane}c). As detailed in reference~~\cite{augustine2023}, selected strip areas with prescribed widths in the range 2-100~$\si{\micro\meter}$ were passivated on the silicon substrate using UV photolithography techniques by means of a thiol-ene click reaction. The actual width of the passivated lines ranged from $9 \pm 1$ to $101 \pm 7~$~\si{\micro\meter} as measured by ellipsometry and ToF SIMS measurements~~\cite{augustine2023}.\\
\subsection*{Debonding set-up}
Swelling-induced debonding of the hydrogel film is achieved by advecting a humid air flow at room temperature on the cooled sample. Under such conditions, the transfer of vapor to the coating was previously shown~\cite{delavoipiere2018} to result from a well-controlled advecto-diffusive mechanism which allows for reproducible and fast (within less than 100~s) saturation of the gel layer. More precisely, we set relative humidity at 60\% and flow rate at 0.6~$\si{L.min^{-1}}$. The water flux to hydrogel coatings was achieved using a homemade setup consisting in a closed Plexiglas chamber equipped with a diverging inlet and converging outlet on opposite sides to obtain a laminar flow of humid gas with controlled rate. In the bottom wall of the chamber, a Peltier module was inserted on which the samples were set. The Peltier module allows to create an homogeneous temperature field at the sample surface and to regulate the sample temperature to a value 10~$\si{\degreeCelsius}$ lower than the ambient temperature. The upper wall of the chamber is a glass window allowing observation of the film by means of a zoom lens (Apo Z16, Leica) and CCD camera (1600 x 1200, 8 bits). Water saturation within the films is first indicated by water condensation droplets on the surface of the PDMA film followed by the formation of a water film at the surface of the gel. In all the experiments to be reported, debonding rates were measured with the films in a fully saturated state from an average of at least three different locations along the patterned lines.\\
%
\section*{Delamination mechanisms} 
\begin{figure} [!ht]
	\centering
	\includegraphics[width=1\linewidth]{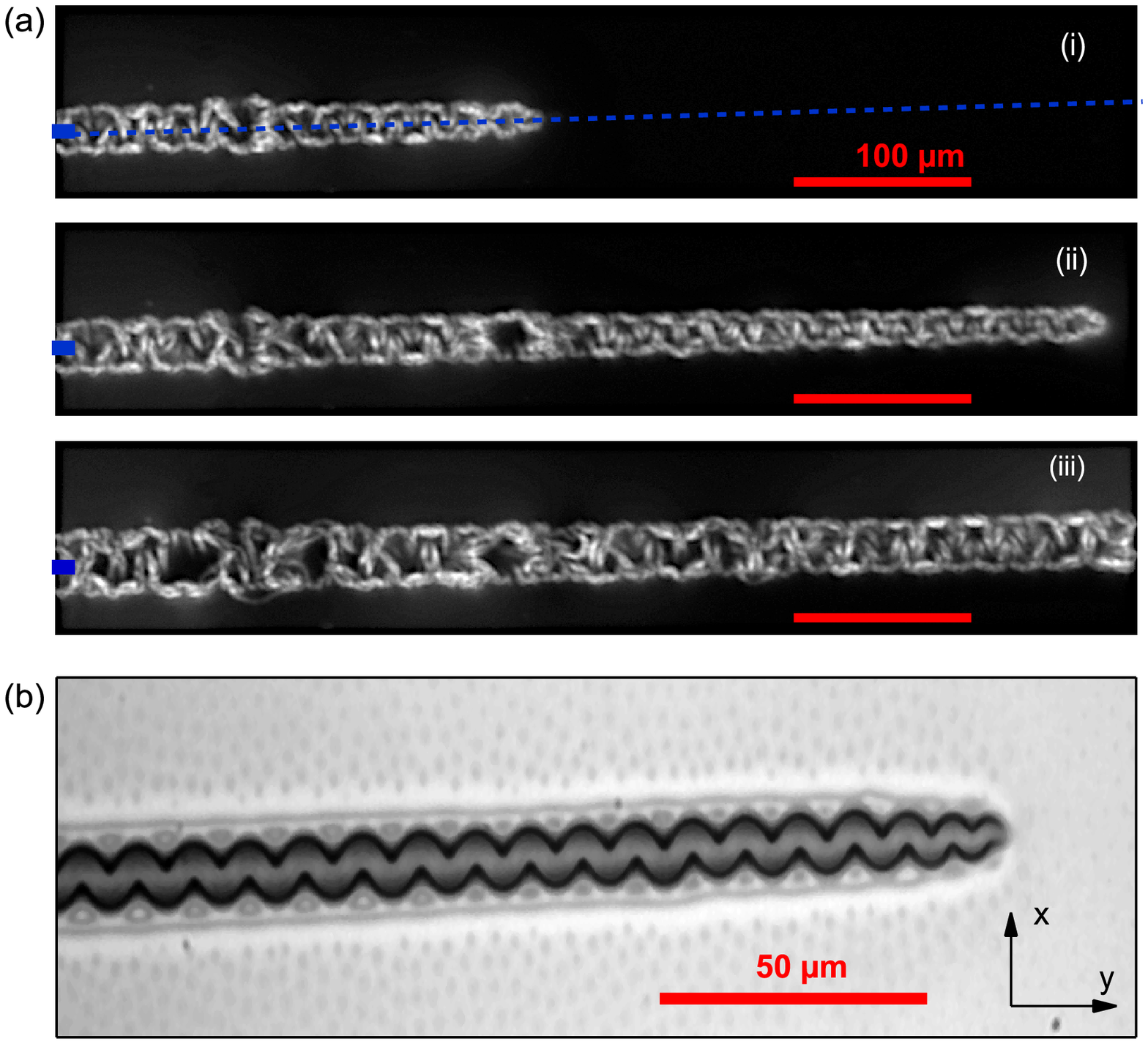}
	\caption{Phone cord debonding mechanisms from a defect line $L_0=9~\si{\micro\meter}$ in width ($h=700~\si{\nano\meter}$, grafting density of 0.6~$\si{\per\square\nano\meter}$, swelling ratio $S=2.7$). (a) images of the swollen film taken (\textit{i}) 360 , (\textit{ii}) 420 and (\textit{iii}) 620~$\si{\second}$ after exposure to humid air. The blue areas and the dotted blue line indicates the line defect width $L_0$ and orientation, respectively. (b) microscopic image of the dried debonded hydrogel with the tip of the debonded area.}
	\label{fig:phone_cord}
\end{figure}

\begin{figure} [!ht]
	\centering
	\includegraphics[width=1\linewidth]{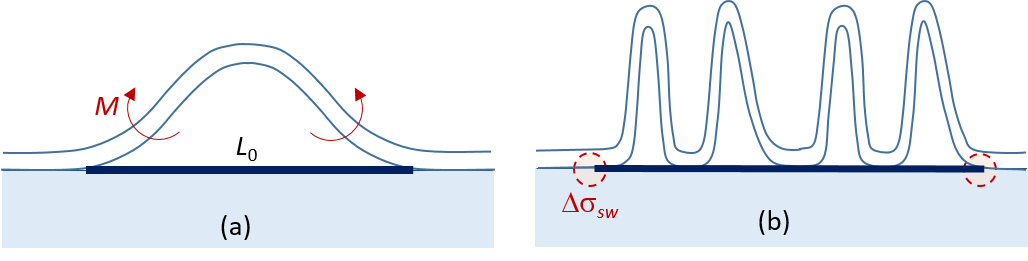}
	\caption{Schematic of potential delamination mechanisms from a defect line of width $L_O$. (a) debonding induced by residual bending stresses in the buckled blister which loads the tip of the interface cracks in a tensile mode; (b) debonding induced by localized differential stress $\Delta \sigma_{sw}$ close to the delamination front. Extensive folding of the film is promoted by the in-plane expansion of the film when it debonds. (b) is the relevant mechanism for swelling induced debonding of soft hydrogel networks since extensive folding of the delaminated film is likely to release crack tip loading by buckling-induced bending stresses.}
	\label{fig:debonding_schematic}
\end{figure}

\begin{figure} [!ht]
	\centering
	\includegraphics[width=1\linewidth]{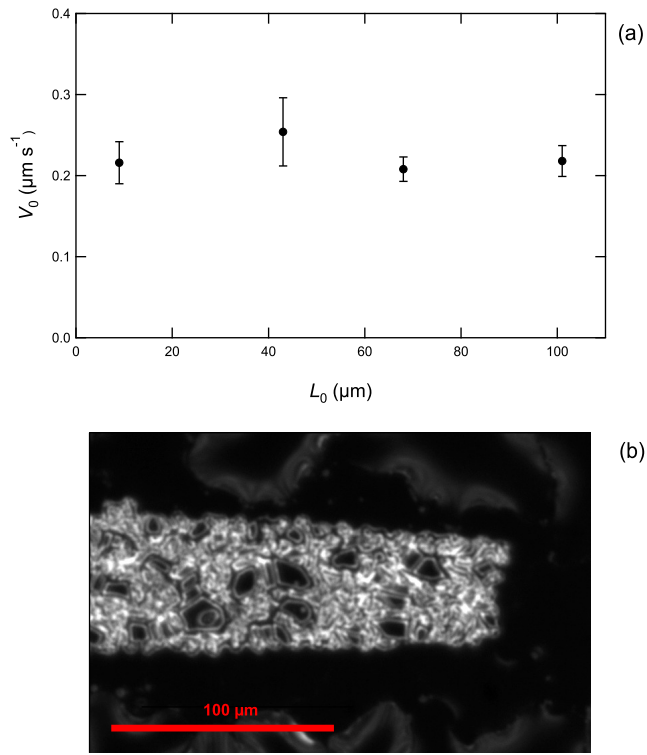}
	\caption{Debonding from defect strips with widths $L_0$ ranging from 9 to 100~\si{\micro\meter} ($h=680$~\si{\nano\meter}, grafting density of 0.6~$\si{\per\square\nano\meter}$, swelling ratio $S=3.8$). (a) initial debonding velocity $V_{0}$ as a function of $L_0$. (b) image of the debonded line for $L_0=100$~\si{\micro\meter} showing extensive folding of the debonded film.}
	\label{fig:v0_L0}
\end{figure}

%
Fig.~\ref{fig:phone_cord}a shows successive images taken during the initial stages of debonding of a film 700~$\si{\nano\meter}$ in thickness from a patterned line defect. The width of the patterned line defect ($L_0=9 \pm 1 \: \si{\micro\meter}$) and its orientation are indicated in the figure by a blue box and dotted line, respectively. Delamination propagates both along the passivated line defect and perpendicular to it, \textit{i.e.} on the grafted part of the film/substrate interface. The debonding rate along the line defect ($y$-axis in Fig.~\ref{fig:phone_cord}a,b) is typically of the order of 1~$\si{\micro\meter\per\second}$ while in the transverse direction ($x$-axis in Fig.~\ref{fig:phone_cord}a,b) it is only in the $10^{-3}\;\si{\micro\meter\per\second}$ range or less for the considered grafting density (0.6~$\si{\per\square\nano\meter}$). Indeed, this difference is consistent with the expectation of a reduced interface strength along the passivated line defect.\\
%
Fig.~\ref{fig:phone_cord}b shows post mortem observations of the tip of the delaminated strip for which the crack is frozen by drying the film and imaging is made with a microscope with enhanced resolution. The images show that the debonded part of the film exhibits a regular folding pattern which looks similar to the so-called telephone-cord architectures reported by Takahashi~\textit{et al.}~\cite{Takahashi2019} and Xu~\textit{et al.}~\cite{xu2016} for various kinds of thin, water swollen, hydrogel films on patterned substrates. Briefly, these features were attributed to buckling instabilities resulting from the swelling induced compressive stresses acting over selected areas of the substrate where adhesion was purposely reduced. As detailed in a theoretical stability analysis of straight-sided blisters by Audoli~\textit{et al.} (the situation depicted Fig.~\ref{fig:debonding_schematic}a), the formation of such telephone cord instabilities in bi-axially compressed films can be accounted for by the residual compressive stress which remains along the longitudinal direction after Euler buckling~\cite{audoly1999,audoly2002}. This secondary buckling is driven by the ratio of the compressive stress $\sigma$ to the buckling stress $\sigma_c$ of a clamped-clamped wide plate. Here, $\sigma_c \propto E(h/L_0)^2$ where $L_0$ and $h$ are respectively the width of the blister and the film thickness. As in our experiments $\sigma \approx E$ (see below), the ratio $\sigma_c/\sigma$ thus scales as $(h/L_0)^2$. Since $h/L_0$ values ranges from $10^{-1}$ to $10^{-2}$, it turns out that $\sigma/\sigma_c >>1$, a situation which was shown to be highly favorable to the formation of telephone cord instabilities.~\cite{audoly1999,audoly2002} In addition to the action of compressive swelling stresses, the buckling of the film is also promoted by the in-plane expansion of the film when it debonds from the substrate as a consequence of the release of the lateral constraints. Accordingly, calculations given below indicate that the stretch ratio along the x-axis is increasing from 1 to 1.9 when the film delaminates.\\ 
As compared to the above mentioned experimental studies, the difference in our experiments is that debonding and buckling over the passivated line defects are followed by the progressive extension of delaminated areas over the \textit{grafted} film/substrate interface, as shown by the lateral extension of the delaminated defect. The driving mechanisms for delamination have been widely addressed in the context of rigid film/substrate systems where the film is in a state of biaxial compression in the unbuckled state as a result, for example, of thermal expansion mismatches. Interface failure in these systems often results in the formation of regions, such as the phone cord blisters, where the film has been buckled away from the substrate in order to minimize its total energy since bending energy is much lower than compressive one.~\cite{hutchinson1992,gioia1997,moon2002} However, when the film buckles away from the substrate, non zero stress intensity factors are induced at the edges of the buckle as a result of the bending moment which promotes further interfacial crack propagation, as schematically depicted in Fig.~\ref{fig:debonding_schematic}a.\\
In our experiments, we noted that delamination also propagates from the free edges of the film specimens with a velocity similar to that achieved along the transverse direction of strip defects. As an example, the debonding velocity from the edge of a film 900~$\si{\nano\meter}$ in thickness and a swelling ratio $S=2.9$ was found to be $0.008 \pm 0.004$~$\si{\micro\meter\per\second}$ while the transverse velocity from a strip defect on the same film was $0.014 \pm 0.004$~$\si{\micro\meter\per\second}$. Such an observation clearly precludes the hypothesis of buckling stresses as the driving force for delamination in our film system. In addition, we also observed that the initial debonding velocity $V_{0}$ in the transverse direction ($x$) is independent on the width $L_0$ of the passivated strip defect\footnote{In these experiments, the swelling ratio of the film is $S=3.8$, which results in values of $V_0$ about two orders of magnitude higher than for a swelling ratio of 2.7, probably as a result of increased swelling stresses.} (Fig.~\ref{fig:v0_L0}a) while the morphology of the debonded film is strongly altered when $L_0$ increases: as shown in Fig.~\ref{fig:v0_L0}b, the shape of the delaminated film is evolving from the single telephone-cord pattern to much more complex morphologies characterized by extensive folding when $L_0$ is increased from 9 to 100~$\si{\micro\meter}$. As schematically depicted in Fig.~\ref{fig:debonding_schematic}b, extensive folding of the delaminated film is likely to relax crack-tip loading by buckling-induced stresses. Here again, the hypothesis of delamination induced by buckling stresses can be discarded from the independence of the debonding velocity on the morphology of the delaminated blisters.\\
In what follows, we will rather make the hypothesis that debonding results from localized differential swelling stresses which arise in the crack tip region as a result of differences in the swelling state between the grafted and delaminated parts of the film.  This hypothesis will be discussed in the light of experiments carried out for a defect width $L_0=9$~$\si{\micro\meter}$, where debonding is initiated from well controlled telephone-cord patterns.\\
%
\subsection*{Time-dependence of the debonded width }
\begin{figure} [!ht]
	\centering
	\includegraphics[width=0.7\linewidth]{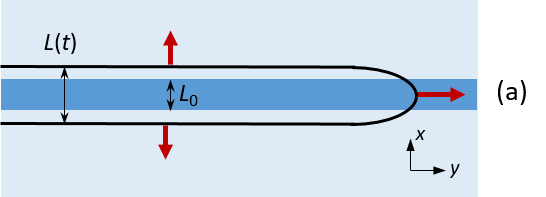}
	\includegraphics[width=1\linewidth]{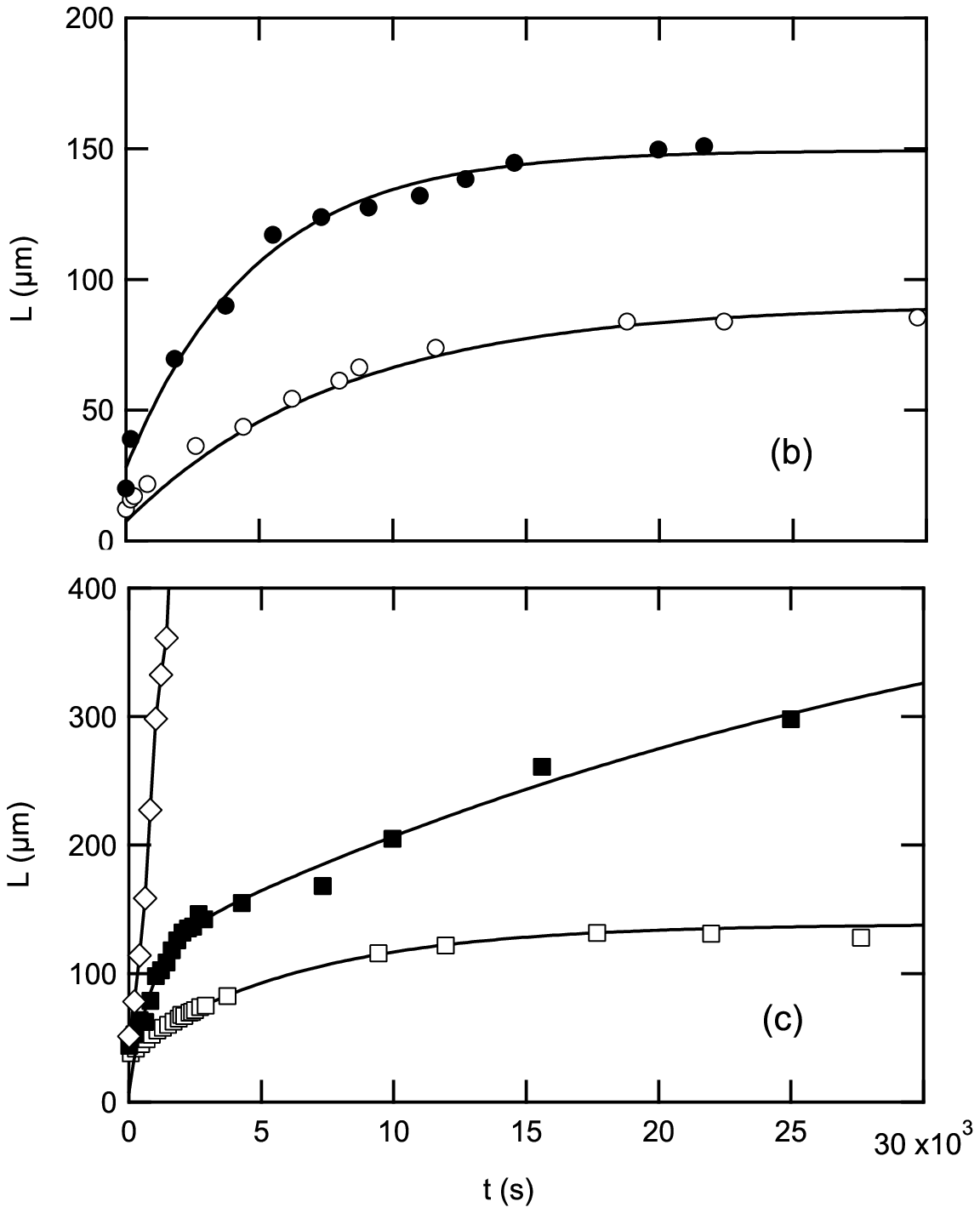}
	\caption{Width $L$ of the debonded strip as a function of time ($L_0=9$~\si{\micro\meter}). (a) schematic of delamination width $L(t)$ along $x$ direction from an interface defect width $L_0$ with $L(t=0) \approx L_0$; (b) film thickness $h=$ ($\circ$) 770 and ($\bullet$) 1130~\si{\nano\meter} with a grafting density of 0.6~$\si{\per\square\nano\meter}$; (c) film thickness $h=$ ($\square$) 480, ($\blacksquare$) 700 and ($\lozenge$) 1060~\si{\nano\meter} with a grafting density of 0.2~$\si{\per\square\nano\meter}$. Solid lines are guides for the eye.}
	\label{fig:l_t}
\end{figure}

\begin{figure} [!ht]
	\centering
	\includegraphics[width=0.9\linewidth]{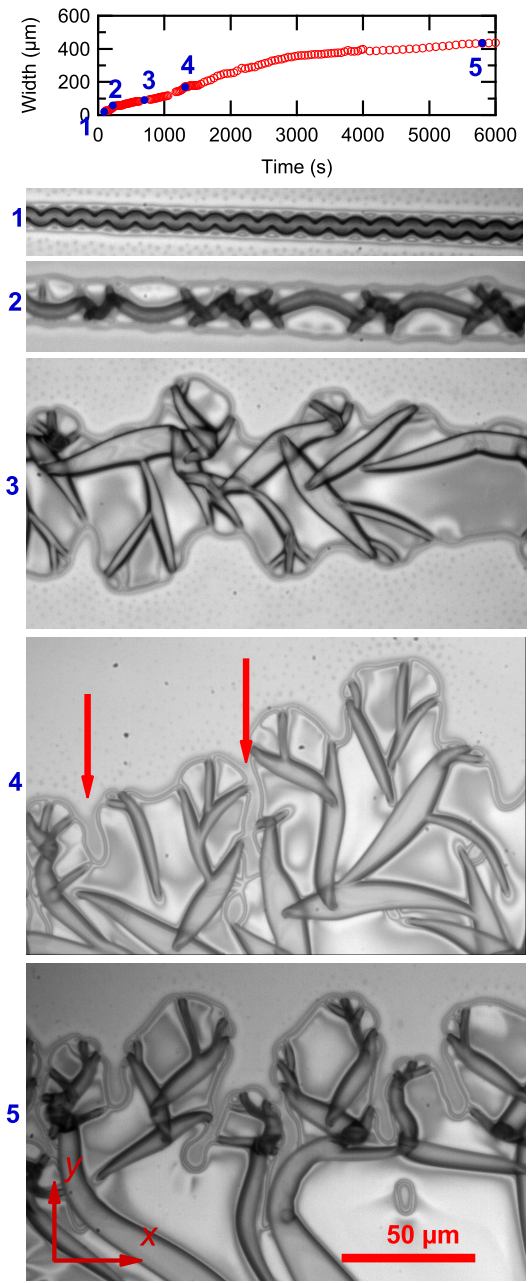}
	\caption{Microscope observation of a dried debonded film at various stages of the delamination process ($h=1.2$~\si{\micro\meter}, grafting density of 0.6~$\si{\per\square\nano\meter}$). Top: width of the debonded line as a function of time. Bottom: images of the debonded lime taken at different times denoted 1 to 5 in the top figure. Only half of the debonded line is shown in (4) and (5). Adhering film fingers with a width about 5 times the swollen film thickness are shown by arrows. The wavelength of the fingering instabilities is about 10 times the swollen thickness.}
	\label{fig:debonding_line}
\end{figure}

\begin{figure} [!ht]
	\centering
	\includegraphics[width=0.9\linewidth]{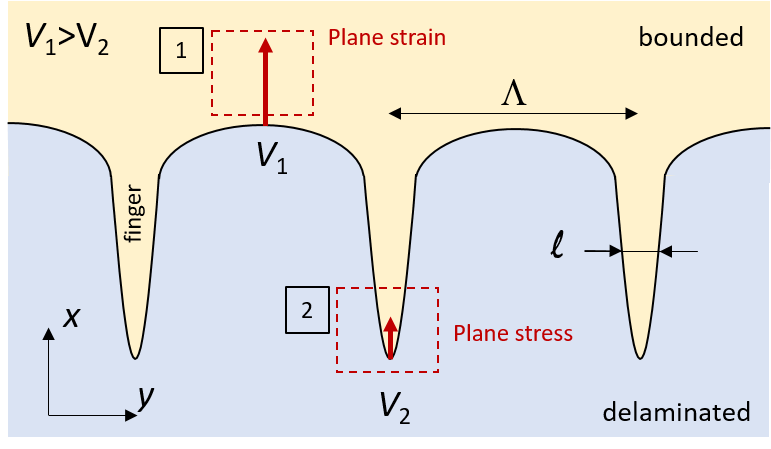}
	\caption{Schematic of debonding along a wavy crack front (blue and yellow areas correspond to the delaminated and bounded part of the film, respectively) corresponding to the situation depicted in fig.~\ref{fig:debonding_line}~(4). In the fingers, the material is assumed to be in a plane stress state as a result of the release of the swelling stress along the $y$ direction. Conversely, the material close to a nearly straight crack front is under a plane strain state. An estimate of the free energy density in both conditions indicates that crack propagation is favored in plane strain (debonding rate $V_1>V_2$).}
	\label{fig:wavy_front}
\end{figure}

We now focus on the time-dependence of the lateral extension $L(t)$ of the delaminated strip for silicon substrates with grafting density of 0.2 and 0.6~$\si{\per\square\nano\meter}$ with film thicknesses $h$ in the range 700 to 1130~\si{\nano\meter}. In fig.~\ref{fig:l_t}, the width of the selected delaminated part is measured over time. It turns out that delamination progressively stops at some debonding width $L$ which increases with film thickness, except for the thickest film with the lower grafting density ($h>480$~\si{\nano\meter}, grafting density 0.2~$\si{\per\square\nano\meter}$) where no delamination arrest is evidenced (diamonds in fig.~\ref{fig:l_t}c).\\
%
Post-mortem microscope observation of the interface cracks frozen at various stages of the debonding process by drying the film revealed that the decrease in the delamination velocity $dL/dt$ is correlated to a progressive change in the shape of the crack line. As an example, fig.~\ref{fig:debonding_line} shows  images of a film 1.2~\si{\micro\meter} in thickness with a grafting density of 0.6~$\si{\per\square\nano\meter}$ together with the corresponding $L(t)$ relationship. It turns out that the decrease in the delamination velocity is associated with the development of a wavy crack line with an approximately constant wavelength $\Lambda$ of about 40~\si{\micro\meter}, \textit{i.e.} about 10 times the thickness of the swollen film. This ratio was preserved when the film thickness was varied between 470 and 1100~\si{\nano\meter} and the grafting density from 0.2 to 0.6~$\si{\per\square\nano\meter}$. As the delamination front moves forward, fingers of adhering film with increased lengths are formed (shown by red arrows in fig.~\ref{fig:debonding_line}). Noticeably, the width of these fingers, denoted $l$, is close to the thickness of the swollen films. For such a situation, the compressive swelling stress within the fingers is likely to be relaxed along the longitudinal $y$-axis, thereby taking the driving mechanism for crack propagation below the threshold stress.\\
%
The development of such a waviness in the delamination front could thus be accounted for by differences in the stress state of the bounded film close to a straight crack front and in the finger-like instabilities. As schematically depicted in fig.~\ref{fig:wavy_front}, a wavy crack front exhibits two distinct areas. In area 1, far from the fingers, the characteristic length in the $y$ direction ($\Lambda$) is much larger than the length scales in the other two directions, \textit{i.e.} $\Lambda>>h$. Conversely, at the finger tip (area 2), the width $l$ of the finger is smaller or comparable with the characteristic lengths in the other two directions {\it i.e.} $h \approx l$. This means that the material close to the nearly straight crack front in between fingers (area 1) is rather in a plane strain state while at the finger tip (area 2), it is under a plane stress condition. Calculations detailed in appendix B indicate that the free energy density of the swollen film under plane stress is lower than that under plane strain which could account for the velocity difference between $V_1$ and $V_2$.\\
A prediction of the wavelength of the instability would require a more quantitative analysis which is beyond the scope of this paper. However, one could anticipate that this wavelength should be proportional to the film thickness which is the only relevant length scale in the problem. Interestingly, experimental~\cite{ghatak2000} and theoretical~\cite{vilmin2010,monch2001} investigations of the peeling of thin films in confined geometries revealed the occurrence of similar fingering instabilities with a characteristic wavelength of the order of 3-4 times the film thickness, a value close to that measured in the present study.\\
%
%
\subsection*{Threshold thickness for delamination }
\begin{figure} [!ht]
	\centering
	\includegraphics[width=1\linewidth]{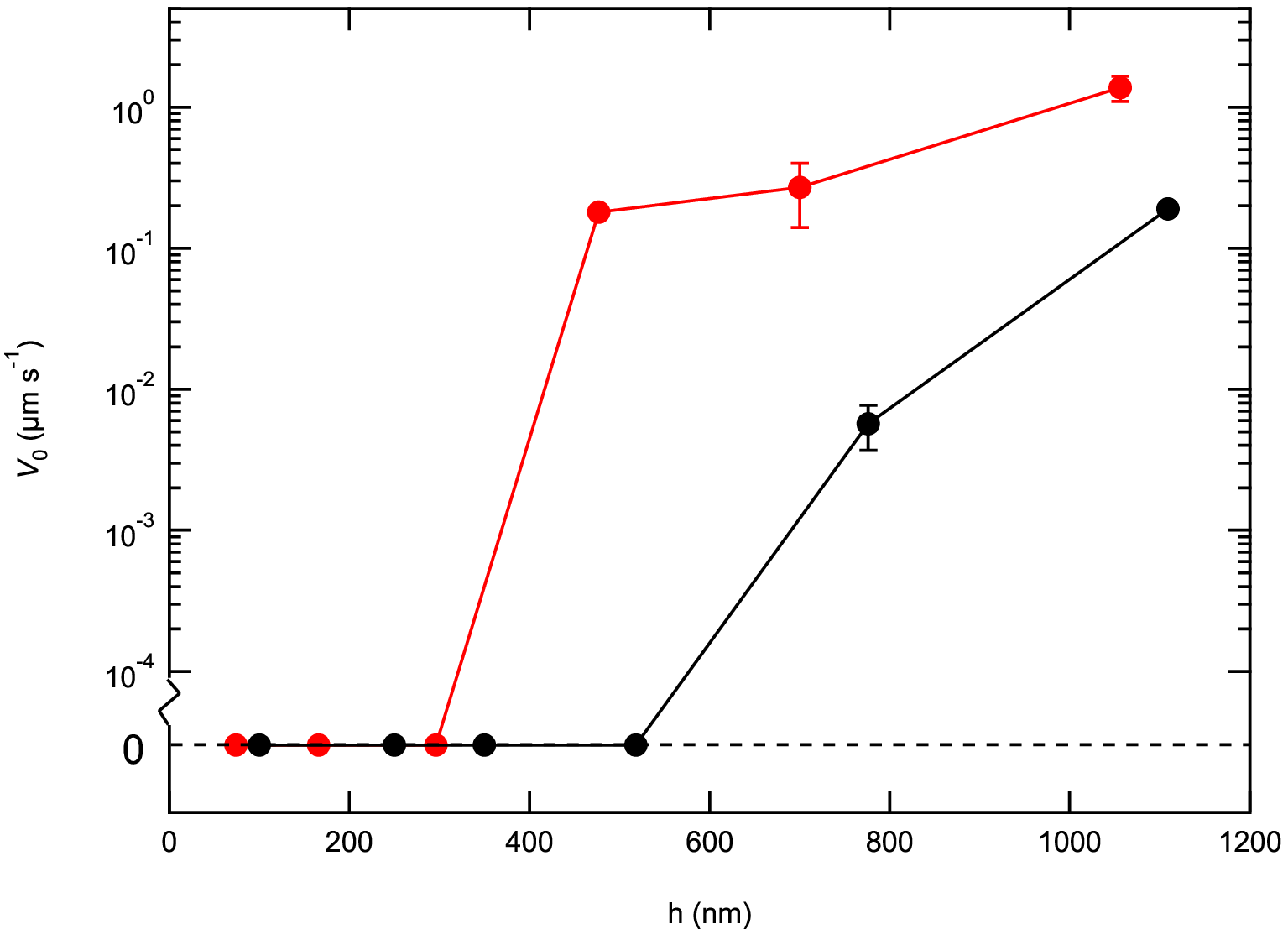}
	\caption{Debonding velocity $V_0$ as a function of dry film thickness $h$ for films with a grafting density of  (\textcolor{red}{$\bullet$}) 0.2~$\si{\per\square\nano\meter}$ and ($\bullet$) 0.6~$\si{\per\square\nano\meter}$.}
	\label{fig:vo_h0}
\end{figure}

%
Fig.~\ref{fig:vo_h0} shows how the initial debonding velocity $V_0$ is varying as a function of the film thickness and grafting density. Here, $V_0$ was measured for a delaminated width of $L=20~\si{\micro\meter}$ (the defect width is  $L_0=9$~\si{\micro\meter}), \textit{i.e.} in a regime where the crack line is still predominantly a straight line as in fig.~\ref{fig:debonding_line}~(1). It turns out that delamination occurs only above a thickness threshold $h^*$ which is decreasing with the grafting density ($h^*=390 \pm 90~\si{\nano\meter}$ and $h^*=650 \pm 130~\si{\nano\meter}$ for 0.2 and 0.6~$\si{\per\square\nano\meter}$, respectively). Above this threshold, orders of magnitude changes in the debonding velocity are measured when the film thickness is increased up to the \si{\micro\meter} range (fig.~\ref{fig:vo_h0}).\\
The existence of a thickness threshold for debonding and the dependence of delamination rate on film thickness can be qualitatively accounted for by a simple scaling argument where the elastic energy $U_e$ per unit length released in the delamination of the constrained film is equated to the fracture energy of the interface $\Gamma_0$. Taking $U_e=\sigma^2h/E$, with $E$ the Young's modulus, the threshold swelling stress for the occurrence of debonding scales as $\sigma_0 \approx \sqrt{\Gamma_0 E/h}$, i.e. as $1/\sqrt{h}$ (as detailed below, the swelling stress is independent on film thickness $h$).\\
It should be noted that the concept that the strain energy release rate increases in proportion of the film thickness and can only propagate an interfacial crack when a critical thickness is exceeded is well established in the literature dealing with hard coating. As an example, the scaling of blister size with film thickness has been used as a tool to measure the adhesion energy in Diamond-Like Carbon (DLC) films based on the estimate of the elastic energy stored within the buckles~\cite{matuda1981,iyer1995,nir1984}. \\
%
In the next section, we develop a refined fracture mechanics analysis of delamination which takes into account the non linearities of the swollen polymer network and makes the assumption that debonding is induced by localized differential swelling stresses instead of buckling stresses. From this analysis, we will derive quantitative values of the interface fracture energy $\Gamma_0$ from the measured thickness threshold $h^*$.\\
%
\section*{Theoretical crack model}
%
In what follows, a nonlinear fracture mechanics approach to the debonding threshold is developed on the basis of a calculation of the amount of energy which is released when the bonded, constrained layer, separates from the substrate and then swells further. Here, this analysis is developed in the limit of a straight crack line propagating along the direction denoted as $x$ in Fig.~\ref{fig:delaminated_film}, \textit{i.e.} for the initial stages of delamination. The main assumption of the model is that debonding arises from localized stresses induced by the differential swelling of the film in the grafted and delaminated states. Specifically, we make the assumption that close to the crack tip, the delaminated film is allowed to swell only in directions $x$ and $z$, i.e. the film is in a plane strain deformation state, while the bonded part the film is only allowed to swell in direction $z$.\\
\subsection*{Swelling  equilibrium}
\begin{figure} [!ht]
	\centering
	\includegraphics[width=1\columnwidth]{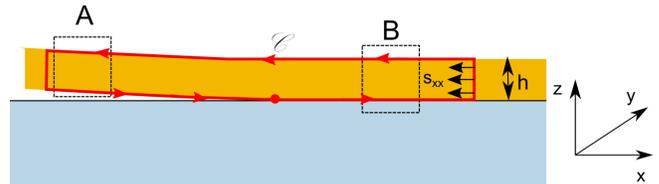}
	\caption{Reference configuration of the delaminated gel film of thickness $h$ on a rigid substrate (before swelling). In zone $A$, the material undergoes free swelling in plane strain, whereas in zone $B$ material is constrained to swell only in direction $z$. By virtue of the plane strain assumption, the gel film at $B$ is under in-plane bi-axial stress. $\mathcal{C}$ is the contour (red line) over which the J-integral in Eq.~(\ref{eq:j_integral}) is evaluated.}
	\label{fig:delaminated_film}
\end{figure}

\begin{figure} [!ht]
	\centering
	\includegraphics[width=0.9\columnwidth]{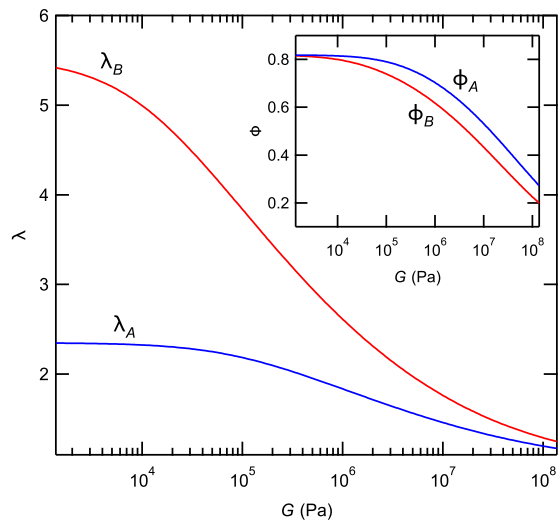}
	\caption{Theoretical swelling ratios $\lambda$ calculated for the bonded (red, $\lambda_B$, Eq.~(\ref{eq:equ_1D})) and the plane strain delaminated parts of the film (blue, $\lambda_A$, Eq.~(\ref{eq:equ_2D})) as a function of the shear modulus $G$ of the drained network ($\chi=0.57$, $kT/\nu=1.38\:10^8\:\si{\pascal}$). Inset: corresponding volume fractions $\Phi_A$ and $\Phi_B$ of water within the swollen network.}
	\label{fig:lambda}
\end{figure}

As a starting point, we first focus on the equilibrium swelling state in the delaminated and bonded parts of the film denoted as $A$ and $B$ in fig.~\ref{fig:delaminated_film}. As detailed by Treloar~\cite{treloar2005}, the swelling equilibrium of the gel network in these two swelling states can be derived from a balance of the free energy of mixing, $W_m$, and the free energy associated to the stretching of network, $W_e$. Following Flory and Rehner~\cite{flory1943}, we take for the free energy of mixing
%
\begin{equation}
	W_{m}=-\frac{kT}{v}\left[vC \ln\left(1+\frac{1}{vC}\right)+\frac{\chi}{1+vC}\right],
	\label{eq:W_mixing}
\end{equation}
where $\nu C$ is the volume of the solvent molecules with respect to the volume of the dry polymer network (\textit{i.e.}, $\nu C=\phi/(1-\phi)$ with $\phi$ the water volume fraction within the swollen gel system) and $\chi$ is a dimensionless measure of the enthalpy of mixing known as the Flory parameter.\\
For the free energy due to the stretching of the polymer network, we used the following expression~\cite{flory1961,wall1951}
\begin{equation}
	W_{e}=\frac{G}{2}\left(\lambda_x^2+\lambda_z^2+\lambda_y^2-3-2 \ln \lambda_x\lambda_z\lambda_y\right),
	\label{eq:W_elastic}
\end{equation}
where $\lambda_{i=x,y,z}$ are the stretch ratios and $G$ is the shear modulus of the drained network. In the above equation, the logarithmic term accounts for swelling (this term would be vanishing for an incompressible rubber, \textit{i.e.} when $\lambda_x \lambda_z \lambda_y=1)$.\\
For one dimensional swelling of the bonded film, $\lambda_z=\lambda_B$, $\lambda_x=\lambda_y=1$ and molecular incompressibility yields $1+\nu C=\lambda_B$. Using equations (\ref{eq:W_mixing}) and (\ref{eq:W_elastic}), the swelling equilibrium is given by
\begin{equation}
	\frac{kT}{v}\left[\ln\left(1-\frac{1}{\lambda_B}\right)+\frac{1}{\lambda_B}+\frac{\chi}{\lambda_B^2}\right]=-G \left(\lambda_B-\frac{1}{\lambda_B}\right).
	\label{eq:equ_1D}
\end{equation}
For plane strain swelling of the debonded film, $\lambda_x=\lambda_z=\lambda_A$, $\lambda_y=1$ and $1+\nu C=\lambda_A^2$ which yields for equilibrium
\begin{equation}
	\frac{kT}{v}\left[\ln\left(1-\frac{1}{\lambda_A^2}\right)+\frac{1}{\lambda_A^2}+\frac{\chi}{\lambda_A^4}\right]=-G \left(1-\frac{1}{\lambda_A^2}\right).
	\label{eq:equ_2D}
\end{equation}
In Fig.~\ref{fig:lambda}, the theoretical values of the swelling ratio in the bonded and delaminated parts of the films are plotted as a function of $G $ for a value of the interaction parameter, $\chi=0.57$ which corresponds to the experimental value of the synthesized PDMA network~\cite{delavoipiere2018}. Although the volume fraction of water $\phi$ within the swollen network (see inset) is, as expected, greater in the delaminated zone ($\phi_A>\phi_B$), it turns out that the release of the lateral constraints results in $\lambda_A<\lambda_B$: the thickness of the debonded film ($A$) is smaller than that of the bounded film ($B$). As mentioned above, the swelling ratio is found to be $\lambda_B=2.7$ from ellipsometry measurements carried out in water. According to eqn~(\ref{eq:equ_1D}), this swelling ratio corresponds to a shear modulus of the drained network  $G=800~\si{\kilo\pascal}$~\footnote{It should be noted that the dry PDMA network is glassy at room temperature. The obtained value of $G$ is thus to be related to that which would be measured above $T_g$ (\textit{i.e.} about 110\si{\degreeCelsius}).}. The calculated thickness of the debonded film is decreased by a factor 1.4 while it is expanding in the direction $x$ by a factor 1.9 with respect to the grafted film (which corresponds to a net increase in the water volume fraction from 0.63 to 0.71). In the bonded part of the film, the true compressive stress (\textit{i.e.} the stress applied to the current, swollen, material configuration) can be written as
\begin{equation}
	\sigma_{xx}=\sigma_{yy}=G \left(\lambda_B-\frac{1}{\lambda_B}\right).
\end{equation}
With $\lambda_B=2.7$, $\sigma_{xx}=1.9~\si{\mega\pascal}\approx 2.4 G$.\\
%
\subsection*{Threshold energy release rate $\Gamma$ for crack extension}
\begin{figure}
	\centering
	\includegraphics[width=1\columnwidth]{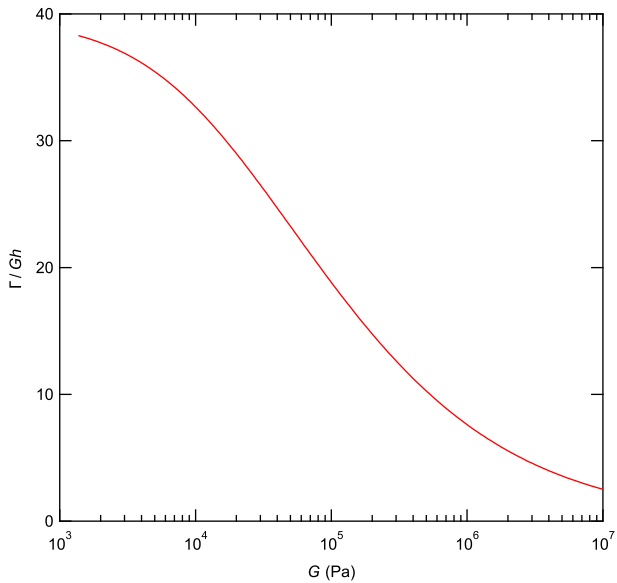}
	\caption{Non dimensional strain energy release rate $\Gamma/Gh$ for the delamination of a film with dry thickness $h$ as a function of the shear modulus $G$ of the drained hydrogel network ($\chi=0.57$, $kT/\nu=1.37\:10^8~\si{\pascal}$).}
	\label{fig:gamma}
\end{figure}

In what follows, we derive the energy $\Gamma$ available for unit crack extension in the reference (\textit{i.e.} dry) configuration using a non linear fracture mechanics approach. Following the work by Bouklas~\textit{et al.}~\cite{bouklas2015} for fracture in hydrogels, we use for that purpose a modified form of the path-independent $J$-integral approach developed by Rice~\cite{rice1968}. Accordingly, $\Gamma$ is evaluated under equilibrium conditions (\textit{i.e.} the chemical potential of the solvent is constant) using the following contour integral $J$ where the usual elastic strain energy density is replaced by the Grand potential $\widehat{W}$
\begin{equation}
	\Gamma\equiv J=\int_\mathcal{C} \left( \widehat{W}_B(\mathbf{F}_B,\mu)N_1-s_{iJ}N_J\frac{\partial x_i}{\partial X_1}\right)dS
	\label{eq:j_integral}
\end{equation}
where according to Eqn 2.24 in~\cite{bouklas2015} $X$ and $x$ denote the coordinates in the reference and deformed configuration, respectively, $\mathbf{F}$ is the deformation gradient tensor, $N_J$ is the $J$ component of outward unit vector normal on the boundary of the reference configuration and $s_{iJ}$ is the nominal stress. $\widehat{W}=W-\mu C$ (with $W=W_m+W_e$) is the Grand free energy density in which the independent variables are the deformation gradient tensor $\mathbf{F}$ and the chemical potential $\mu$ of the solvent molecules~\cite{hong2009}.\\
Evaluation of the $J$-integral along the contour $\mathcal{C}$ (Fig.~\ref{fig:delaminated_film}) yields
\begin{equation}
	\Gamma=h \left[ \left(  \widehat{W}_B(\mathbf{F}_B,\mu) - s_{xx}(B)\right)- \left (\widehat{W}_A(\mathbf{F}_A,\mu) \right) \right]
\end{equation}
where the subscripts $A$ and $B$ denote that the quantity $\widehat{W}$ is evaluated at $A$ and $B$ (see fig.~\ref{fig:delaminated_film}). Here we have considered that the gel is in a state of equilibrium and that the chemical potential of the solvent molecules within the gel is homogeneous and equal to that of the solvent outside the network. In the above expression, we also made the assumption that the only relevant stress in the bonded part of the gel is the compressive swelling stress $s_{xx}$ and that the delaminated part of the gel film is traction free. The latter is justified by the extensive folding of the debonded film we observed experimentally. After some calculations detailed in Appendix A, the strain energy release rate writes
\begin{multline}
	\frac{\Gamma}{Gh}=\left[\frac{\lambda_B^2-1-2\ln\lambda_B}{2}+ 2 \ln \lambda_A+\frac{1}{\lambda^2_A}-\frac{1}{\lambda_B} + \lambda_B-1\right] \\ +\frac{kT}{\nu G}\left[ \left(2\chi-1\right)\left(\frac{1}{\lambda^2_A}-\frac{1}{\lambda_B}\right)+\frac{\chi}{\lambda^2_B}-\frac{\chi}{\lambda^4_A}\right].
	\label{eq:release_rate}
\end{multline}
We used the results in fig.\ref{fig:lambda} to obtain $\lambda_A$ and $\lambda_B$ to calculate the normalized strain energy release rate $\Gamma/Gh$ which is reported in fig.\ref{fig:gamma} as a function of the shear modulus $G$ of the network for a value of the Flory interaction parameter $\chi=0.57$.\\
%
%
\section*{Discussion and conclusion}
%
Swelling stresses within the investigated PDMA hydrogel films were observed to induce delamination from pre-existing strip defects with low adhesion. The velocity and the extent of debonding depend on both film thickness and the amount of reactive thiol groups grafted on the silicon substrate prior to film deposition. A threshold thickness $h^*$ for the initiation of debonding was especially evidenced. This threshold thickness corresponds to a situation where the strain energy release rate  reaches the fracture energy of the interface, $\Gamma=\Gamma_0$. Taking $G=800$~\si{\kilo\pascal}, we can determine $\lambda_A$ and $\lambda_B$ from fig.~\ref{fig:lambda} and use this value to calculate the non dimensional strain energy release rate $\Gamma/Gh$ from Eqn~\ref{eq:release_rate}. We find $\Gamma/Gh=8.4$ which means that at the onset of debonding $\Gamma_0=8.4 G h^*$. For the measured values of $h^*$ (from $390$ to $650$~\si{\nano\meter}), this value corresponds to a threshold fracture energy which ranges from $\Gamma_0$=2.6 to 4.3~\si{\joule\per\square\meter} when the thiol density is increased from $\Sigma$=0.2 to 0.6~$\si{\per\square\nano\meter}$. \\
For a comparison, an independent estimate of the fracture energy can be derived from simple molecular arguments taking into account the energy which is dissipated in the breaking of a stretched network chain grafted onto the rigid silicon substrate. For that purpose, we follow an argument initially proposed by Lake and Thomas~\cite{lake1967} for rubber fracture which was recently extended by Akagi~\textit{et al.}~\cite{akagi2013} to the molecular description of fracture energy within water swollen polymer gels. Basically, this theory postulates that when any of the chain bonds breaks, the total bond energy of each bond in the chain is irreversibly lost. Assuming that only chains grafted onto the silicon substrate break, the threshold energy for crack initiation is 
\begin{equation}
	\Gamma_0=\eta U_b \Sigma
	\label{eq:gamma_0}
\end{equation}
where $\Sigma$ is the areal density of chains grafted on the silicon wafer, $U_b$ is the dissociation energy of a carbon-sulfur bond ($\approx 4 \times 10^{-19} \:\si{\joule}$) and $\eta$ is the average number of monomers of the chains. From the value of the shear modulus, $G=800$~\si{\kilo\pascal}, we can estimate $\eta$ as the average monomer number between crosslinks: $\eta=\rho RT/Gm_m\approx 30$ with $m_m$=99~$\si{\gram\per\mole}$ the monomer mass and $\rho$ the density. We assume here that the surface density of the grafted chains and of the thiol groups are equal. According to eqn.~\ref{eq:gamma_0}, for $\Sigma$=0.1 to 1~\si{\per\square\nano\meter}, the threshold fracture energy is found to be of the order of $\Gamma_0\approx$1 to 10~\si{\joule\per\square\meter} which is consistent with the values derived from eqn~(\ref{eq:release_rate}).\\
The order of magnitude agreement between the threshold fracture energy derived from our model and a Lake and Thomas estimate of the energy required to break covalent bonds at the interface tends to support our assumption of a vanishing poroelastic dissipation close to the crack tip at the onset of crack propagation. The validation of our fracture model from experiments would, however, deserve further investigations, especially regarding the effects of changes in the swelling ratio as a result of varying network crosslink densities. From Eq.~\ref{eq:release_rate}, we expect the threshold thickness $h^*$ to depend non linearly on the associated changes in the shear modulus and swelling ratio.\\
Experimental results show that above the crack threshold, order of magnitude changes in the initial debonding rate $V_0$ are also induced when the thickness or the grafting density are varied (Fig.~\ref{fig:vo_h0}). A theoretical analysis of $V_0$ would require much more involved calculations taking into account the dissipative processes at skate when the crack front is advancing. As detailed by Bouklas~\textit{et al.},~\cite{bouklas2015} the determination of the energy release rate in such situations requires to take into account solvent diffusion at the crack tip, \textit{i.e.} to separate the energy lost in diffusion from the enregy available to drive the crack growth. Such an approach would require extensive numerical simulations which are beyond the scope of this study.\\
%
\section*{Acknowledgments}
This project has received funding from the European Union’s Horizon 2020 research and innovation program under the Marie Skłodowska-Curie grant agreement No 754387. It was also granted by Saint Gobain Research Paris. C.-Y. Hui and B. Zhu are supported by NSF CMMI grant no.-1903308. Part of this work was carried out on the Platform for Nano-Characterization (PFNC) and was supported by the “Recherche Technologique de Base” program of the French National Research Agency (ANR). One of us (A. Augustine) is indebted to Ekkachai Martwong for his kind support during the synthesis of the hydrogels systems. C.-Y. Hui also thank ESPCI Paris for hosting his visit.\\
%
\appendix
%
\section*{Appendix}
\numberwithin{equation}{subsection}
\renewcommand{\thesubsection}{\Alph{subsection}}
\subsection{Derivation of the strain energy release rate $\Gamma$}
In the reference (\textit{i.e.} dry) configuration the energy $\Gamma$ available for unit crack extension is derived  using the following J-integral expression
\begin{equation}
	\Gamma=h \left[ \left(  \widehat{W}_B(\mathbf{F}_B,\mu) - s_{xx}(B)\right)- \left (\widehat{W}_A(\mathbf{F}_A,\mu) \right) \right]
	\label{eq:gamma}
\end{equation}
%
According to Hong~{et al.}~\cite{hong2008}, the grand free energy can be written as
\begin{eqnarray}
	\widehat{W}(\mathbf{F},\mu)=\frac{G}{2}\left(I-3-2\ln J \right) \\
	\nonumber	
	 -\frac{kT}{\nu}\left[ \left(J-1\right)\ln \frac{J}{J-1}+\frac{\chi}{J}\right] \\ 
	\nonumber
	-\frac{\mu}{\nu}\left(J-1\right)
	\label{eq:W_}
\end{eqnarray}
where $J=\det \mathbf{F}$ and $I=\Tr( \mathbf{F}^T \mathbf{F})$. Using the Flory-Rehner model~\cite{flory1943} one can shown that, in plane strain
\begin{eqnarray}
	\widehat{W}_A(\mathbf{F}_A,\mu)=G \left(\lambda_A^2-1-\ln \lambda_A^2 \right)\\
	\nonumber	
	-\frac{kT}{\nu}\left[ \left(\lambda_A^2-1\right)\ln \frac{\lambda_A^2}{\lambda_A^2-1}+\frac{\chi}{\lambda_A^2}\right] \\
	\nonumber
	 -\frac{\mu}{\nu}\left(\lambda_A^2-1\right)	
	\label{eq:W_A}
\end{eqnarray}
with
\begin{equation}
	\mathbf{F}_A=\begin{vmatrix} \lambda_A & 0 & 0 \\ 0 & 1 & 0  \\ 0 & 0 & \lambda_A \end{vmatrix}
	\label{eq:FA}
\end{equation}
and
\begin{eqnarray}
	\widehat{W}_B(\mathbf{F}_B,\mu)=\frac{G}{2}\left(\lambda_B^2-1-\ln \lambda_B^2 \right) \\
	\nonumber
	-\frac{kT}{\nu}\left[ \left(\lambda_B-1\right)\ln \frac{\lambda_B}{\lambda_B-1}+\frac{\chi}{\lambda_B}\right] \\
	\nonumber
	 -\frac{\mu}{\nu}\left(\lambda_B-1\right)
	\label{eq:W_B}
\end{eqnarray}
with
\begin{equation}
	\mathbf{F}_B=\begin{vmatrix} 1 & 0 & 0 \\ 0 & 1 & 0  \\ 0 & 0 & \lambda_B \end{vmatrix},
\end{equation}
where the relationships between the equilibrium swelling ratios $\lambda_A$,$\lambda_B$ and the chemical potential of the swollen gel $\mu$ is given by
\begin{equation}
	\frac{\mu_A}{\nu}=G \left(1-\frac{1}{\lambda_A^2}\right)-\frac{kT}{\nu} \left[  \ln\frac{\lambda_A^2}{\lambda_A^2-1}-\frac{1}{\lambda_A^2}-\frac{\chi}{\lambda_A^4}\right]
	\label{eq:mu_A}
\end{equation}
\begin{equation}
	\frac{\mu_B}{\nu}=G \left(\lambda_B-\frac{1}{\lambda_B}\right)-\frac{kT}{\nu} \left[\ln \frac{\lambda_B}{\lambda_B-1}-\frac{1}{\lambda_B}-\frac{\chi}{\lambda_B^2}\right]
	\label{eq:mu_B}
\end{equation}
with $\mu_A=\mu_B$. The free energy at $A$ can be expressed in term of $\lambda_A$ only by substituting (\ref{eq:mu_A}) into (\ref{eq:W_A}) to eliminate the chemical potential,\textit{i.e.}
\begin{eqnarray}
	\widehat{W}_A(\mathbf{F}_A,\mu)=G \left(\frac{\lambda_A^2-1}{\lambda_A^2}-\ln \lambda_A^2\right)\\
	\nonumber
	-\frac{kT}{\nu}\left[ 1+\frac{2 \chi -1}{\lambda_A^2}-\frac{\chi}{\lambda_A^4}\right]  
	\label{eq:W_A1}
\end{eqnarray}
In the same way, substituting (\ref{eq:mu_B}) into (\ref{eq:W_B}) and taking $s_{xx}(B)=\partial W_e/\partial \lambda_B$, it comes
\begin{eqnarray}
	\widehat{W}_B(\mathbf{F}_B,\mu)-s_{xx}(B)=\frac{G}{2} \left(\lambda_B^2-1-2 \ln \lambda_B\right)\\
	\nonumber
	+G\left(\frac{\lambda_B^2-1}{\lambda_B}\right)	-\frac{kT}{\nu}\left[1+ \frac{2 \chi -1}{\lambda_B}-\frac{\chi}{\lambda_B^2}\right] 
	\label{eq:W_B1}
\end{eqnarray}
From Eq.~\ref{eq:gamma}, we obtain Eq.~\ref{eq:release_rate} that gives $\Gamma$ as a function of $\lambda_A$ and $\lambda_B$.
%
\numberwithin{equation}{subsection}
\renewcommand{\thesubsection}{\Alph{subsection}}
\subsection{Estimate of the free energy density in the finger-like instabilities}
In order to estimate the free energy density of the adhering film in the finger-like instabilities, we assume that the swelling stresses are largely relaxed due to the release of lateral constraints when $\ell \approx h$ (area 2 in fig.~\ref{fig:debonding_line}). Such a situation is approximated by considering that the film within the finger is under a plane stress condition equivalent to isotropic swelling. The corresponding free energy $W_iso$ is calculated from Eq.~\ref{eq:W_} with

\begin{equation}
	\mathbf{F}_{iso}=\begin{vmatrix} \lambda_{iso} & 0 & 0 \\ 0 & \lambda_{iso} & 0  \\ 0 & 0 & \lambda_{iso} \end{vmatrix},
\end{equation}

where the stretch ratio $\lambda_{iso}$ is obtained from the Flory Rehner theory~\cite{flory1943} as 

\begin{equation}
	\frac{kT}{v}\left[\ln\left(1-\frac{1}{\lambda_{iso}^3}\right)+\frac{1}{\lambda_{iso}^3}+\frac{\chi}{\lambda_{iso}^6}\right]=-G \left(\frac{1}{\lambda_{iso}}-\frac{1}{\lambda_{iso}^3}\right).
	\label{eq:equ_3D}
\end{equation}
Conversely, when the adhering film is close to a straight crack front (area 1 in fig.~\ref{fig:debonding_line}), we consider that the film is under a plane strain condition with the strain deformation tensor given by Eq.~\ref{eq:FA}. The corresponding free energy density is obtained from Eq.~\ref{eq:W_A}.\\
Taking $G=800 \:\si{\kilo\pascal}$, $kT/\nu =1.37\: 10^{8}~\si{\pascal}$, $\lambda_A=1.87$ and $\lambda_{iso}=1.58$, the difference in the free energy between plane strain $W_{PE}$ and plane stress $W_{PS}$ is $\Delta W=W_{PS}-W_{PE}=6.37\:10^5~\si{\joule\per\cubic\meter}=637~\si{\kilo\pascal} \approx G$ which means that the energy density within the adhering fingers is less than that close to a straight crack front and that the difference is the order of magnitude of the shear modulus.
%
%
\bibliographystyle{rsc}

	\end{document}